\documentclass[12pt]{article}
\pdfoutput=1

\usepackage{amsmath}
\usepackage{amssymb}
\usepackage{amsfonts}
\usepackage{mathrsfs}

\usepackage{mathrsfs}
\usepackage{fullpage}
\usepackage{setspace}
\usepackage{bm}
\usepackage{bbm}

\usepackage[normalem]{ulem}
\usepackage{enumerate}
\usepackage{yfonts}

\usepackage{psfrag}

\usepackage{graphicx}
\usepackage{cancel}
\usepackage{slashed}

\usepackage[font=small,labelfont=bf]{caption}
\usepackage{subcaption}

\usepackage[colorlinks=true]{hyperref} 
\hypersetup{
    bookmarks=true,         
    unicode=false,          
    pdftoolbar=true,        
    pdfmenubar=true,        
    pdffitwindow=false,     
    pdfstartview={FitH},    
    pdftitle={Instantons and Entanglement Entropy},    
    pdfauthor={Arpan Bhattacharyya, Ling-Yan Hung, and Charles M. Melby–Thompson},     
    pdfnewwindow=true,      
    colorlinks=true,       
    linkcolor=blue,          
    citecolor=red,        
    filecolor=magenta,      
    urlcolor=cyan           
} 

\newcommand{\be}{\begin{equation}}
\newcommand{\bea}{\begin{eqnarray}}
\newcommand{\ee}{\end{equation}}
\newcommand{\eea}{\end{eqnarray}}

\newcommand{\bes}{\begin{equation*}}
\newcommand{\ees}{\end{equation*}}
\newcommand{\beas}{\begin{eqnarray*}}
\newcommand{\eeas}{\end{eqnarray*}}

\newcommand{\bmat}{\begin{bmatrix}}
\newcommand{\emat}{\end{bmatrix}}

\begin{document}

\begin{titlepage}

\thispagestyle{empty}

\begin{flushright}
YITP-18-116
\end{flushright}


\numberwithin{equation}{section}
{
\begin{center}

\hfill \\
\hfill \\
\vskip 0.25in

\noindent{{\Large \textbf {Post-Quench Evolution of Complexity and Entanglement  in a Topological System}}}\\
\vspace{1cm}
	
Tibra Ali$^{(a)}$, Arpan Bhattacharyya$^{(b),(c)}$ ,  
S. Shajidul Haque$^{(d),(e)}$,\\ Eugene H. Kim$^{(e)}$, Nathan Moynihan$^{(d)}$, \\ ~~~~\\

$ {}^{(a)}$ {\it Perimeter Institute, \\
\it 31 Caroline Street North\\
 Waterloo, Ontario, Canada, N2L 2Y5 \\
}	
       
       ${}^{(b)}${\it Indian Institute of Technology,
Gandhinagar,\\ \it Gujarat 382355, India }

$ {}^{(c)}$ {\it Center for Gravitational Physics, \\
\it Yukawa Institute for Theoretical Physics (YITP), Kyoto University, \\
\it Kitashirakawa Oiwakecho, Sakyo-ku, Kyoto 606-8502, Japan\\
}	

$ {}^{(d)}$ {\it{The Laboratory for Quantum Gravity \& Strings,\\
	Department of Mathematics \& Applied Mathematics,\\
	University of Cape Town,\\
	Private Bag, Rondebosch, 7701, South Africa}\\ 
}

$ {}^{(e)}$ {\it Department of Physics, University of Windsor, \\
\it 401 Sunset Avenue\\
Windsor, Ontario, Canada, N9B 3P4\\
}	

\end{center}

\begin{abstract}

We investigate the evolution of complexity and entanglement following a quench in a one-dimensional topological system, namely the Su-Schrieffer-Heeger model. We demonstrate that complexity can detect quantum phase transitions and  shows signatures of revivals; this observation provides a practical advantage in information processing.  We also show that the complexity saturates much faster than the entanglement entropy in this system, and we provide a physical argument for this. Finally, we demonstrate that complexity is a less sensitive probe of topological order, compared with
measures of entanglement.
\end{abstract}
\vfill


}
\end{titlepage}
\newpage
\tableofcontents

\section{Introduction}

In recent years, concepts from quantum information (QI) have helped to mold ideas 
in high-energy physics \cite{qithepreview}, namely 
anti de-Sitter/conformal field theory (AdS/CFT) duality \cite{maldacena}. 
Recent progress has made evident that the entanglement of a (boundary) CFT is  related to the 
emergence of the bulk geometry \cite{entanglegravity}.
This relationship becomes even more puzzling in the context of black hole --
entanglement is not sufficient to capture the dual geometric description; certain bulk quantities 
continue to evolve even after the boundary CFT has equilibrated \cite{susskind1}.
As a way out of this conundrum, Susskind proposed that the boundary quantity which continues
to evolve after equilibration is the state's {\sl complexity} \cite{susskind1,susskind2}.

Central in QI are notions of {\sl distance} and {\sl uncertainty} \cite{QITbook}. 
Entanglement is quantified by measures of uncertainty.  
Complexity is a concept from the theory of computation \cite{computation,QITbook} --
it is the shortest distance between some reference state $| \psi_R \rangle$ and a 
target state $| \psi_T \rangle$; operationally, it quantifies the minimal number of operations 
needed to manipulate $| \psi_R \rangle$ to $| \psi_T \rangle$.
Entanglement in quantum field theory has been well studied, leading to a variety
of insights \cite{entanglereview1,jphysa}.
An understanding of complexity in quantum field theory (QFT) is in its infancy --
circuit complexity  using Nielsen's approach \cite{nielsen} was studied for
free scalar QFT in \cite{myers}, for interacting scalar QFT in \cite{My},
and for free fermionic theories in \cite{bangalore,myersfermion}; 
complexity using the Fubini-Study distance was studied for free scalar QFT in \cite{heller}, 
using a path-integral in \cite{path}; 
\cite{TM} discussed thermofield-double states, and 
\cite{brazil,caputaheller, paper1} considered quantum quenches in bosonic QFTs.



In this work,  we investigate the time evolution of complexity and entanglement following a quantum quench ---
we compare the evolution of the complexity (for both the circuit complexity and Fubini-Study approaches) and fidelity with 
measures of entanglement. 
To begin with, we elucidate the physics governing the evolution of these quantities, and in particular, their saturation.
In holography, complexity is expected to saturate later than the entanglement entropy,
but for the system considered here, the opposite occurs. This counterexample raises confusion regarding the 
holographic description of complexity. 
Secondly, we investigate revivals in the complexity and explore whether it can detect different phase transitions.
Finally, we demonstrate that the circuit complexity is not sensitive to the system's {\sl topological order}, 
but this is captured by the Fubini-Study approach and by measures of entanglement.
The circuit complexity is expected to be more sensitive than the Fubini-Study approach \cite{brownsusskind,heller}; 
here, the opposite occurs.
Our medium for these investigations is a one-dimensional topological insulator (TI) \cite{TIreviews}, 
namely the Su-Schrieffer-Heeger (SSH) model \cite{SSH}.
This is a tractable system exhibiting a nontrivial phase diagram, for which 
these ideas can be scrutinized in detail.
It should be noted, so far complexity (and its comparison with entanglement) has only been explored 
in the context of simple toy models; this work takes steps toward bringing complexity closer to physical systems, 
and in particular, experiment. 
Indeed, the SSH model was introduced to understand the electronic properties of polyacetylene \cite{SSH}.
It has been realized experimentally in cold atom systems \cite{bloch1,bloch2,takahashi,gadway}.


\section{Compelxity and Entanglement for SSH model}

The SSH model is described by the Hamiltonian
\begin{equation}
 \hat{H} = \sum_l \left[ 
   q \left( a^{\dagger}_l b^{\phantom \dagger}_l 
          + b^{\dagger}_l a^{\phantom \dagger}_l \right)
 + q' \left( a^{\dagger}_{l+1} b^{\phantom \dagger}_l
           + b^{\dagger}_l a^{\phantom \dagger}_{l+1} \right)
    \right] \ ,
\label{ssh}
\end{equation}
where $a_l$ ($b_l$) destroys a fermion on site-$l$ of sublattice-$A$ ($B$), 
and $q$ ($q'$) is the intracell (intercell) tunneling matrix element. 
%
The topological properties are determined by ($q$, $q'$) -- 
for $|q| < |q'|$ ($|q| > |q'|$), the system is a TI (non-TI); 
$|q| = |q'|$ is a quantum critical point (QCP), described by a $c$=1 CFT.
In what follows, we work in the Neveu-Schwarz sector;
the system is readily analyzed in momentum space -- expanding the fermion operators in Fourier modes
[$k=(2\pi/N)(m+1/2)$ with $1\leq m \leq N$]
\begin{equation}
 a_l = \frac{1}{\sqrt{N}} \sum_k e^{ikl} \tilde{a}_k
 \ {\rm and } \ 
 b_l = \frac{1}{\sqrt{N}} \sum_k e^{ikl} \tilde{b}_k  \ ,
\end{equation}
one obtains
\begin{subequations}
\begin{equation}
 \hat{H} = \sum_k
    ( \tilde{a}^{\dagger}_k, \tilde{b}^{\dagger}_k )
    \left( \begin{array}{cc}
      0 \ & q + q' e^{-ik}  \\
      q + q' e^{ik} \ & 0  
     \end{array} \right)  
    \left( \begin{array}{c}
      \tilde{a}_k \\ \tilde{b}_k
     \end{array} \right) .
\label{sshfourier}
\end{equation}
(\ref{sshfourier}) is diagonalized by a Bogoliubov transformation, details of it are given in the section  (I) of supplementary material. Then we have, 
\begin{equation}
 \hat{H} = \sum_k E_k \left( 
     \alpha^{\dagger}_k \alpha^{\phantom \dagger}_k
  - \beta^{\dagger}_k \beta^{\phantom \dagger}_k 
  \right)  \ .
\label{ssheigenmodes}
\end{equation}
\end{subequations}
In Eq.~(\ref{ssheigenmodes}), $E_k = \sqrt{ |\Delta_k| + 2 q q' \cos k}$
with $\Delta_k= q + q' \exp(-ik)$; 
$\alpha_k = (\tilde a_k + \tilde{b}_k~ \Delta_k/|\Delta_k| )/\sqrt{2}$ and 
$\beta_k = (- \tilde{a}_k~ \Delta_k^{*}/|\Delta_k| + \tilde b_k)/\sqrt{2}$ 
destroy a fermion of momentum-$k$ in the conduction and valence band, respectively. For more details please refer to section (\ref{secc3}) of Appendix~(\ref{app}).


We are interested in the quench dynamics of the SSH model -- 
\begin{equation}
 \underline{t<0}: \hat{H} = \hat{H}_< (q^{\phantom '}_<,q'_<)
 \ \  ,  \ \
 \underline{t>0}: \hat{H} = \hat{H}_> (q^{\phantom '}_>,q'_>)
 \ .
\end{equation}
In what follows, we take $| \psi_R \rangle$ to be the ground state of $\hat{H}_<$ and 
$| \psi_T \rangle = \exp(-i t \hat{H}_>) | \psi_R \rangle$.
More explicitly, let $\{ \alpha_{<k}, \beta_{<k} \}$ denote the destruction operators of $\hat{H}_<$.
The initial state has the valence band of $\hat{H}_<$ completely filled 
\begin{subequations}
\begin{equation}
 | \psi_R \rangle = \prod_k \beta^{\dagger}_{<k} | 0 \rangle  \ ,
\label{referencestate}
\end{equation}
where $| 0 \rangle$ is the Fock vacuum. It is one of the three phases described below equation (\ref{ssh}), depending on the choice of the parameter $(q_<,q_<').$
Then after doing a time evolution $ \exp(-i t \hat{H}_>)$ one obtains
\begin{equation}
 | \psi_T \rangle = \prod_k \frac{1}{\sqrt{1 + | z_k(t) |^2 } }
   \left( \beta^{\dagger}_{<k} + z_k(t)~ \alpha^{\dagger}_{<k} \right) 
   | 0 \rangle
\label{targetstate}
\end{equation}
\end{subequations}
where
 \begin{equation}
 \textstyle z_{k}(t)=\frac{|\Delta_k^>||\Delta_k^<| \Big(|\Delta_k^>| \text{Im} (\Delta_k^<) 
      + |\Delta_k^<|\text{Im}(\Delta_k^>)\Big)\sin (E_k^>\, t)}
  {\Delta_k^>\Delta_k^{* <}\Big(|\Delta_k^>||\Delta_k^<|\cos(E_k^>\, t) 
      + i\, \text{Re}(\Delta_k^>)\text{Re}(\Delta_k^<)\sin( E_k^>\, t)\Big)} \ .
 \nonumber
\end{equation}
For more details please refer to the section (\ref{secc}) of Appendix~(\ref{app}). \par  It is common in the literature that, when one computes the circuit complexity one typically chooses a direct product state (e.g.  \cite{myers}). But in our case, we were interested in detecting phase transitions by  circuit complexity, caused by the time evolution of the initial ground state of $H_<(q_<,q_<').$ Hence it is natural to compute circuit complexity between the ground state of initial Hamiltonian $H_<(q_<,q_<')$ and the state after the time evolution, as it is providing us with a new notion of distance between these two states.

We will be computing the circuit complexity for 
the state (\ref{targetstate}) w.r.t the state (\ref{referencestate}). For more details of reference and traget state please refer to section (\ref{secc4}) of Appendix~(\ref{app}).
These states can be characterized by their {\sl correlation matrix}, $\hat{C}_{k}(t)$,
\begin{equation}
 \hat{C}_{k}(t) = \langle \psi(t) | \Psi^{\phantom \dagger}_{k} 
                       \Psi^{\dagger}_{k} | \psi(t) \rangle \ ,
\end{equation}
where $\Psi_{k}^T = ( \tilde{a}_{k}, \tilde{b}_{k} ).$ Details are given in  Appendix~(\ref{app1}).
To obtain the circuit complexity, we evolve the initial correlation matrix 
$\hat{C}_k(t=0)$ by the unitary operator $U_k(s)$ \cite{myersfermion}
\begin{equation}
 \tilde{C}_k(s) \equiv U_k^{\phantom \dagger} (s) 
                       \hat{C}_k(t=0)U_k^{\dagger}(s) 
\end{equation}
with $\tilde{C}_k(s=1) = \hat{C}_k(t)$, and 
$U_k(s)$ is parameterized as the path-ordered exponential
\begin{equation}
 U_k(s) = {\cal P} \exp \left[ - i \sum_I \int_0^1 \hspace{-0.07in} ds~
              Y_k^I(s) M_I \right]  \ ,
\label{gates}
\end{equation}
where the $\{ M_I \}$ are group generators, the $\{ Y_k^I \}$ are control functions,
and $s$ parameterizes the path.
Then, we define a ``cost functional," ${\cal C}[\{U_k(s) \}]$, on the space of unitaries via
\begin{equation}
 {\cal C}[ \{U_k(s) \} ] = \int_0^1 \hspace{-0.07in} ds
    \sqrt{ \sum_k \sum_I  |Y_k^I(s)|^2  } \ .
\label{cost1}
\end{equation}
Eq.~(\ref{cost1}) is minimized w.r.t. the $\{ Y_k^I \}$ as a function of  ``s"; 
this determines the optimal path, i.e. the geodesic  on the space of unitaries.
The circuit complexity is the minimum value of  ${\cal C}[ \{ U_k(s) \} ]$.  For more details please refer to the section ~(\ref{secc1}) of Appendix~ (\ref{app1}).

We will compare the results for the circuit complexity with those obtained from the Fubini-Study line element.
To this end, the space of states in (\ref{targetstate}) is given a Riemannian 
structure \cite{statemanifold} --- 
we start with the Bures distance \cite{bures,uhlmann}
\begin{subequations}
\begin{equation}
 D_{12}^2 = 2 \left[ 1 - {\cal F}_{12} \right] 
\label{bures}
\end{equation}
where ${\cal F}_{12}$ is the {\sl fidelity}
\begin{equation}
  {\cal F}_{12} = \left| \langle \psi_1 | \psi_2 \rangle \right| \ ;
\end{equation}
\end{subequations}
considering two nearby states 
$| \psi(\{ z_k \}) \rangle$ and $| \psi(\{ z_k + dz_k \}) \rangle$, 
we obtain the Fubini-Study line element
\begin{equation}
 ds^2 = \sum_k  \frac{ |dz_k|^2}{ \left( 1 + |z_k|^2 \right)^2}  \ ,
\label{metricCP1}
\end{equation}
where $ds^2 \equiv {\rm lim}_{| \psi_2 \rangle \rightarrow | \psi_1 \rangle} D^2_{12}$.
One recognizes each term in (\ref{metricCP1}) to be the line element for a two-sphere $S^2$ 
in the $CP^1$ representation -- 
the state manifold is $S^2 \times S^2 \times \cdots \times S^2$;
the distance is given by \cite{metricspaces}, 
\begin{subequations}
\begin{equation}
  s = \sqrt{ \sum_k s_k^2 }  \ ,
\label{productspace}
\end{equation}
where 
\begin{equation}
 s_k = \int_{0}^{t} \hspace{-0.07in} dt'~ 
  \frac{1}{1 + |z_k|^2 }  \left| \frac{dz_k}{dt'} \right| \ . 
\label{S2metric}
\end{equation}
\end{subequations}
%
The complexity ${\cal C}$ is the length of the geodesic connecting $| \psi_R \rangle$ and $| \psi_T \rangle$. More details regarding this are given in Appendix~(\ref{app2}).


In what follows, we consider the following quenches: 
(1) TI $\leftrightarrow$ non-TI
(2) non-TI $\leftrightarrow$ QCP
(3) TI $\leftrightarrow$ QCP.
Results are shown for the following sets of parameters: 
(1) TI -- ($q$=0.2, $q'$=1)
(2) non-TI -- ($q$=1, $q'$=0.2)
(3) QCP -- ($q$=1, $q'$=1).
To characterize the quenches, we first consider the system's entanglement.
We partition the total system into two subsystems, $A$ and $B.$  We consider the 
correlation matrix restricted to subsystem-$A$, $\hat{C}^A(t)$,
whose elements are ($m,n \in A$)
\begin{equation}
 \hat{C}^A_{mn}(t) = \frac{1}{N} \sum_k \exp[i k (m-n)]~\hat{C}_k(t) \,.
\end{equation}
The eigenvalues $\{ \lambda_n \}$ of $\hat{C}^A(t)$ are referred to as 
the entanglement spectrum; measures of entanglement are functions of the
entanglement spectrum \cite{peschel,ehk}. More details regarding this are given in the section ~(\ref{secc2}) of  Appendix~(\ref{app1}).

\begin{figure}[t]
\centering
\scalebox{1}{\includegraphics{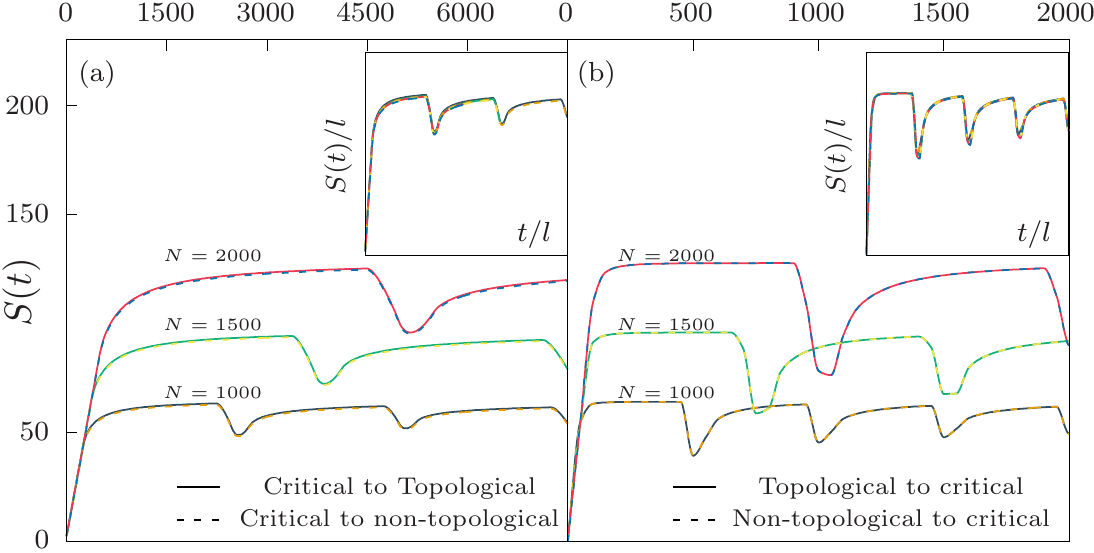}}\\
\vspace{0px}\scalebox{1}{\includegraphics[trim = {0 0 0 0.15cm}]{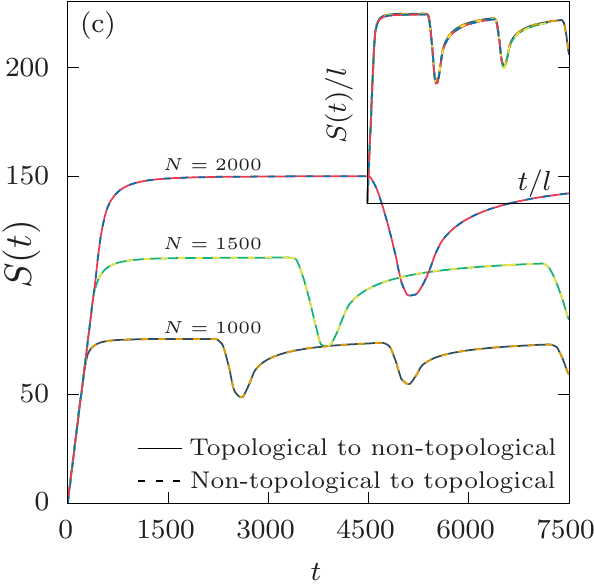}}
\caption{Evolution of EE  for  
$N$=1000, $N$=1500, $N$=2000 and subsystem of 
$l$=100, $l$=150, $l$=200 respectively for  quenching from 
(a) the critical point to a massive phase
(b) a massive phase to the critical point 
(c) between massive phases.
Insets: EE vs time scaled by the partition size, $S(t)/l$ vs $t/l$.
}
\label{fig:entanglement}
\end{figure}

Fig.~(\ref{fig:entanglement}) shows results for the entanglement entropy (EE)
\begin{equation}
 S = - \sum \left[ \lambda_n \ln \lambda_n
   + (1 - \lambda_n) \ln ( 1 - \lambda_n ) \right]  \ .
\end{equation}
The EE first grows linearly and then saturates.
At later times, one sees revivals due to the finite system size. 
Furthermore, we see that the amplitude is larger when quenching between 
massive phases (compared to when quenching to/from the QCP).
The (pseudo-) period of the revivals is larger when quenching to a massive phase 
(compared to when quenching to the QCP).
As shown in the insets, by working with scaled variables, namely $S/l$ vs. $t/l$, the results
collapse onto universal curves.
While the results for the quenches have similar features, there are quantitative differences --
the EE is (slightly) larger in the TI phase, as the TI has greater entanglement.

\begin{figure}[th!]
\centering
\scalebox{1.2}{\includegraphics{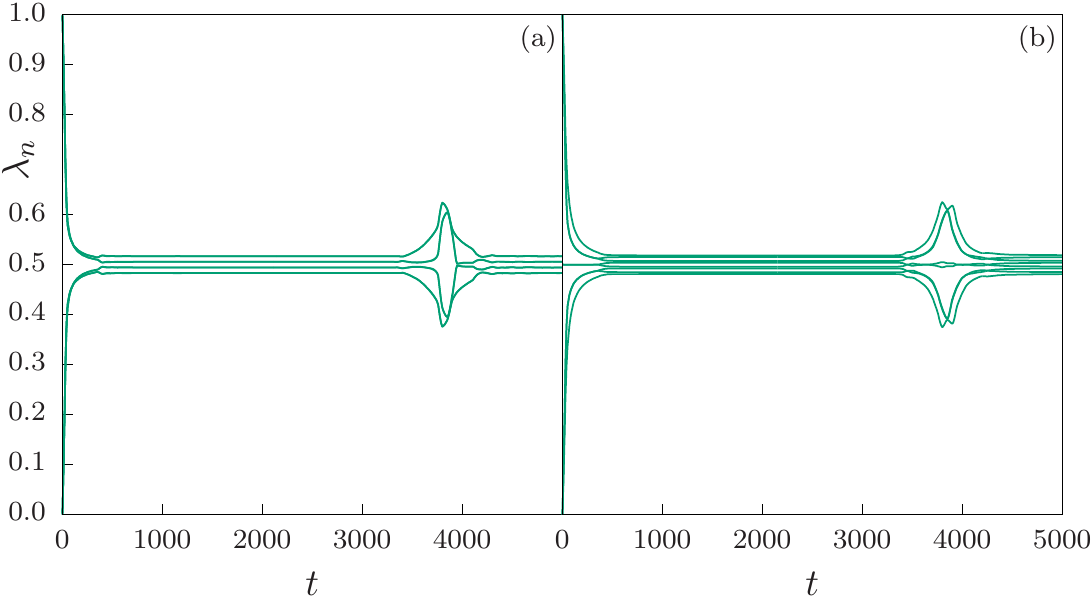} }
\caption{Evolution of entanglement spectrum $N$=1500 and subsystem of $l$=150 for 
quenching from 
(a) the non-topological to the topological phase
(b) the topological to the non-topological phase. 
}
\label{fig:ES}
\end{figure}
To further elucidate the system's properties, the entanglement spectrum is shown in Fig.~(\ref{fig:ES}) --
Fig.~(\ref{fig:ES}a) (Fig.~(\ref{fig:ES}b)) shows the entanglement spectrum for a quench from the non-TI (TI) 
to the TI (non-TI) phase. 
This shows very distinctly the differences in the quenches --  
the TI has ``zero-modes" in the entanglement spectrum, namely states where $\lambda_n \simeq 1/2$.
These are due to the edge states that are present only in the TI phase \cite{wenbook,ehk}.  
When one starts initially in the TI phase, the zero-modes stay pinned.
[Results for the other quenches are qualitatively similar, namely quenches to/from the TI 
are clearly seen in the evolution of zero-modes \cite{scirep,british}.]


\begin{figure}[h!]
\centering
\scalebox{0.70}{\includegraphics{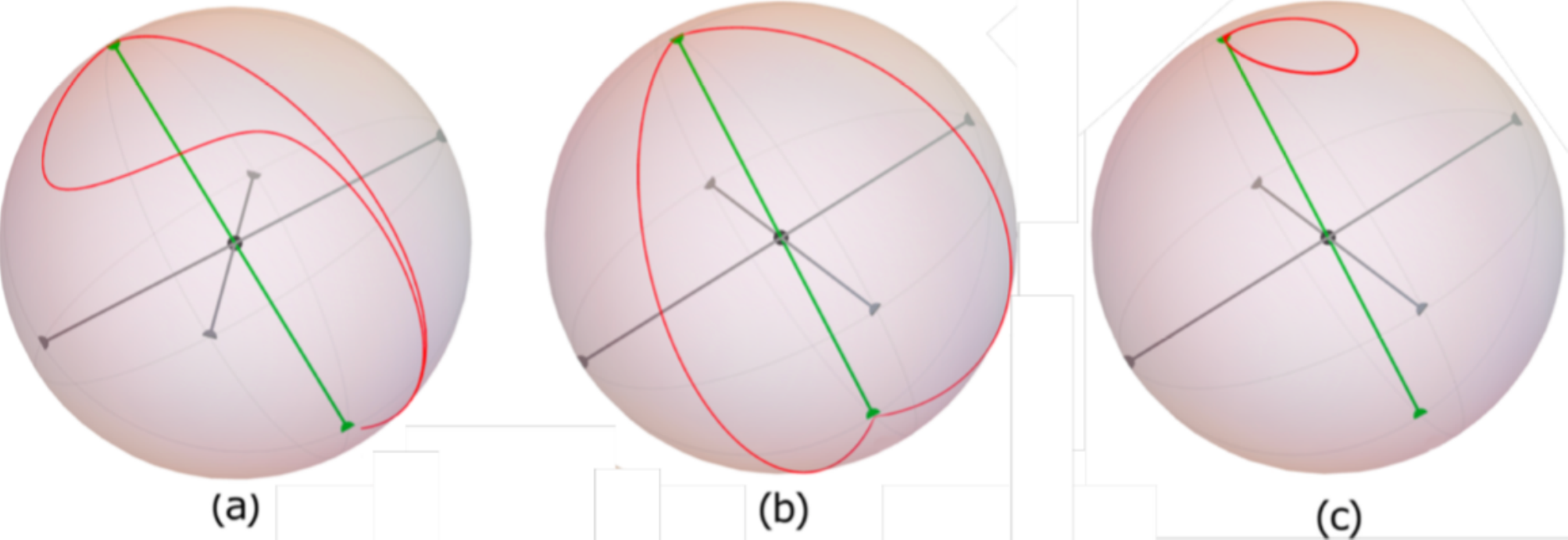} }
\caption{Motion on the Bloch sphere for transition from 
the non-TI to QCP phase ($k=(2\pi/N)(m+1/2)$):
(a) $m$ = 470 (b) $m$ = 505 (c) $m$ = 870  for $N$=1000.}
\label{fig:bloch}
\end{figure}
We now discuss the system's complexity.
Fig.~(\ref{fig:bloch}) shows the evolution of the target state for a transition from the non-TI
to the QCP for several values of $k$.
As discussed above, (\ref{targetstate}) describes a trajectory on $S^2$ for each $k$.
We see that Hamiltonian evolution gives rise to a variety of behaviors -- depending on the value
of $k$, the evolution can be close to the geodesic for some values (Figs.~(\ref{fig:bloch}a) and (b)), 
while it can be rather distant for other values (Fig.~(\ref{fig:bloch}c)).
Hence, for some values of $k$ the path from Hamiltonian evolution gives the complexity, 
while for other values of $k$ the contribution to the complexity comes from a very different path. 
[Similar results are obtained for the other quenches.]

\begin{figure}[th!]
  \centering
\includegraphics[trim=1.5cm .5cm 1cm .2cm,scale=0.60]{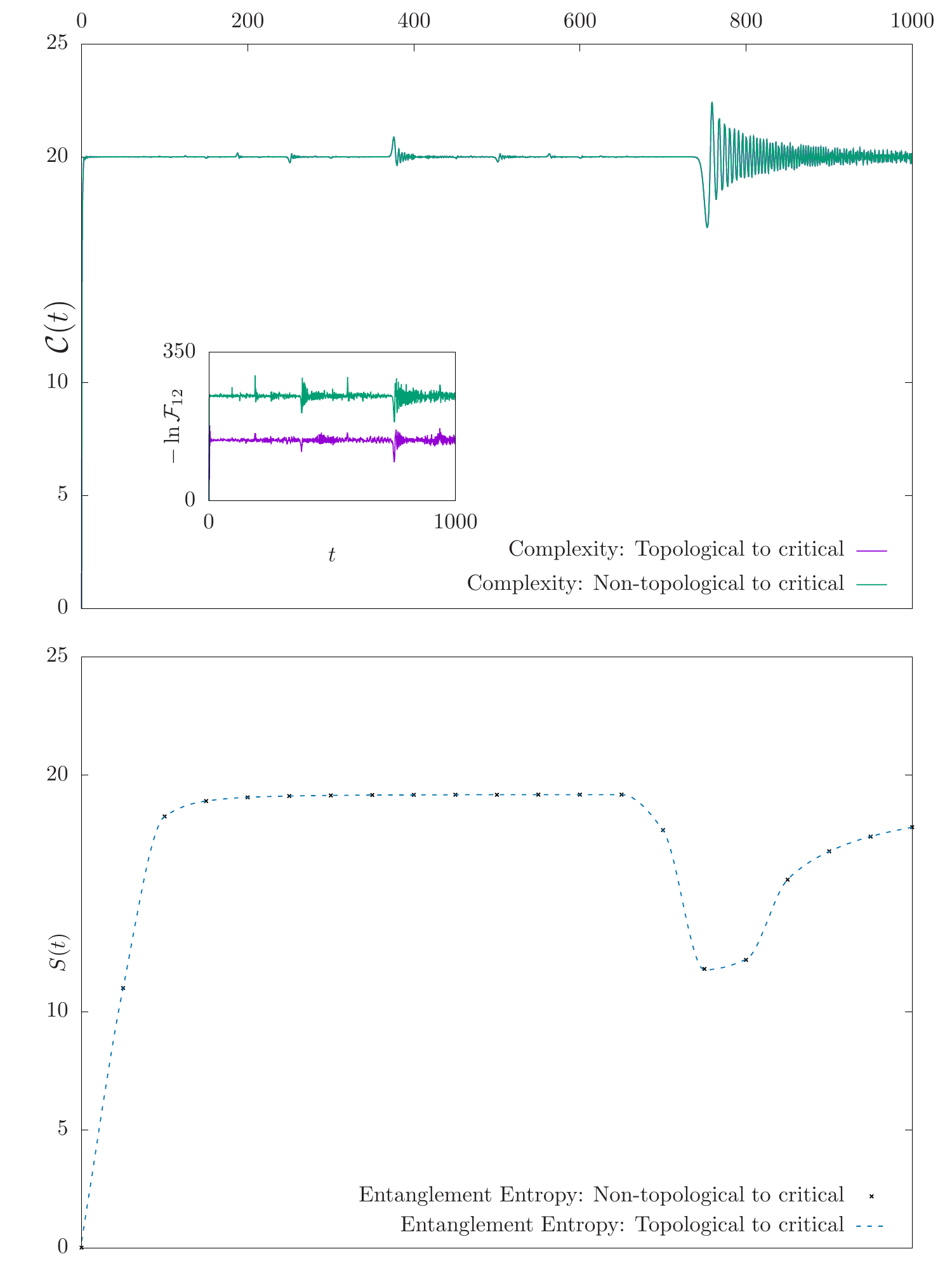}~~~~~~~~~~~~\includegraphics[trim=1.5cm .5cm 1cm .2cm,scale=0.60]{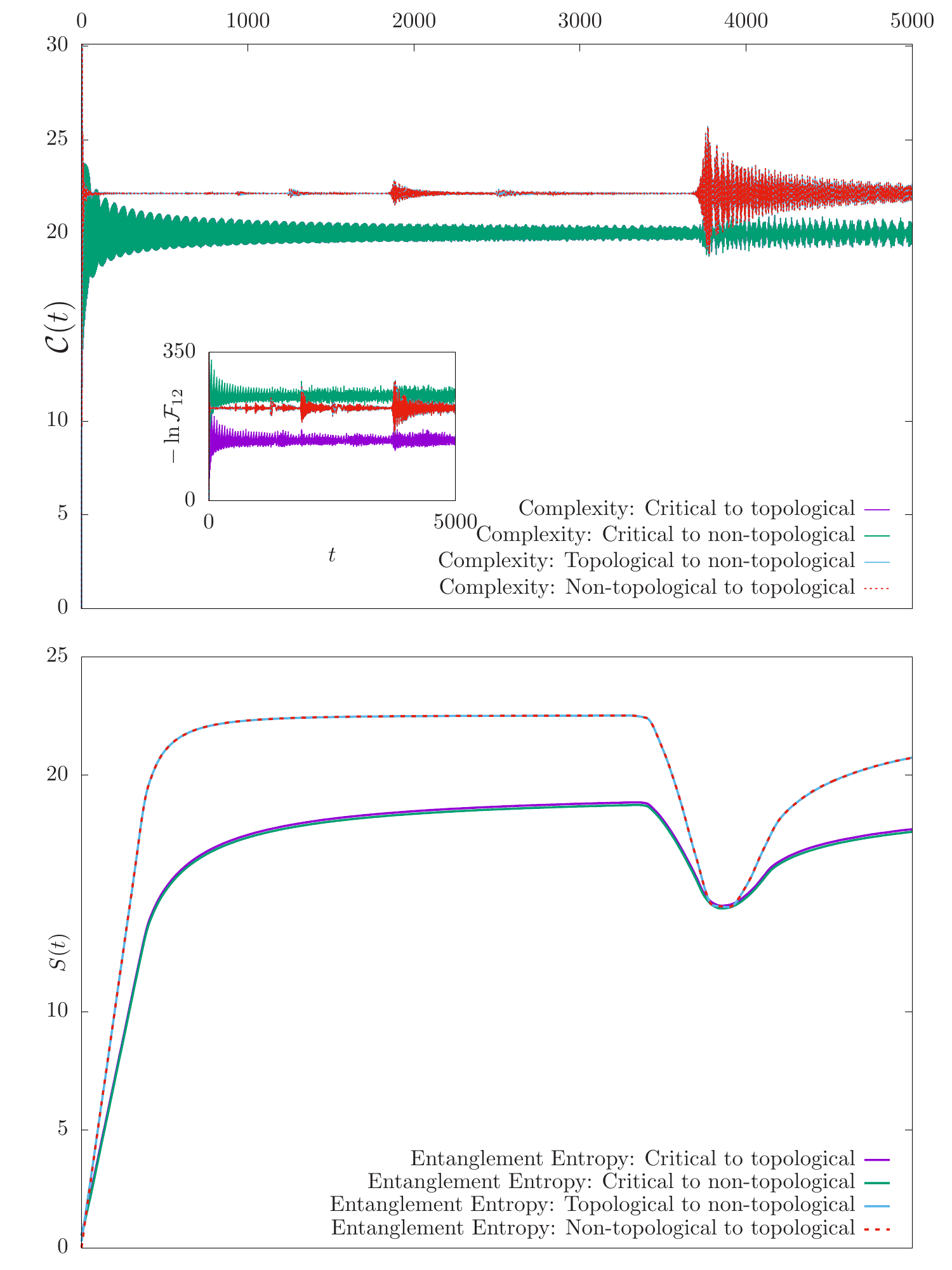}
  \caption{The circuit complexity for $N$=1500. 
Quenching 
(a) from a massive phase to the critical point
(b) to a massive phase.
Insets: Negative logarithm of the fidelity: $-\ln {\cal F}_{12}$.
}
\label{fig:circuitcomplexity}
\end{figure}
\newpage
\begin{figure}[th!]
	\centering
	\includegraphics[trim=1.5cm .5cm 1cm .2cm,scale=0.60]{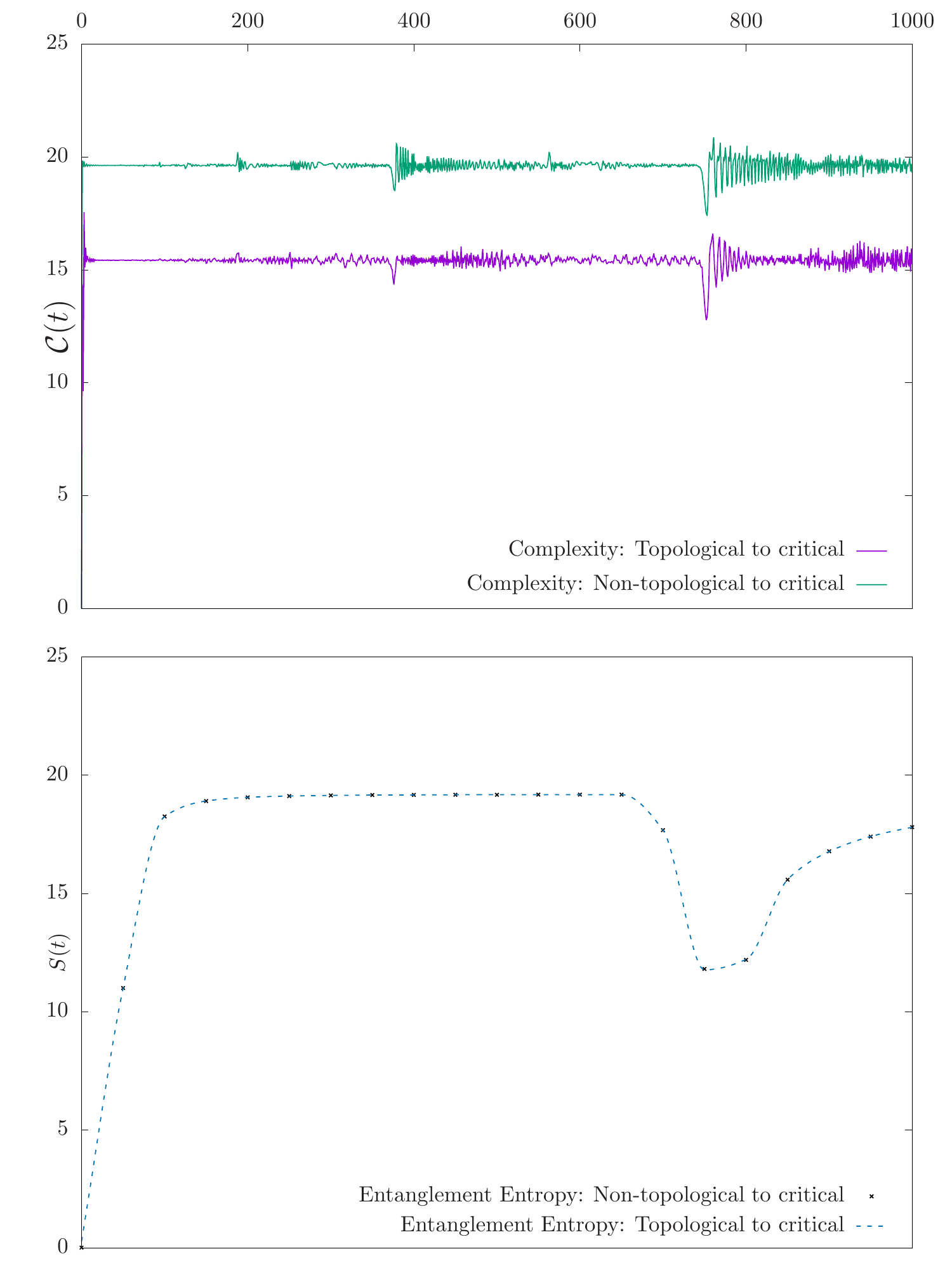}~~~~~~~~~~~~
	\includegraphics[trim=1.5cm .5cm 1cm .2cm,scale=0.60]{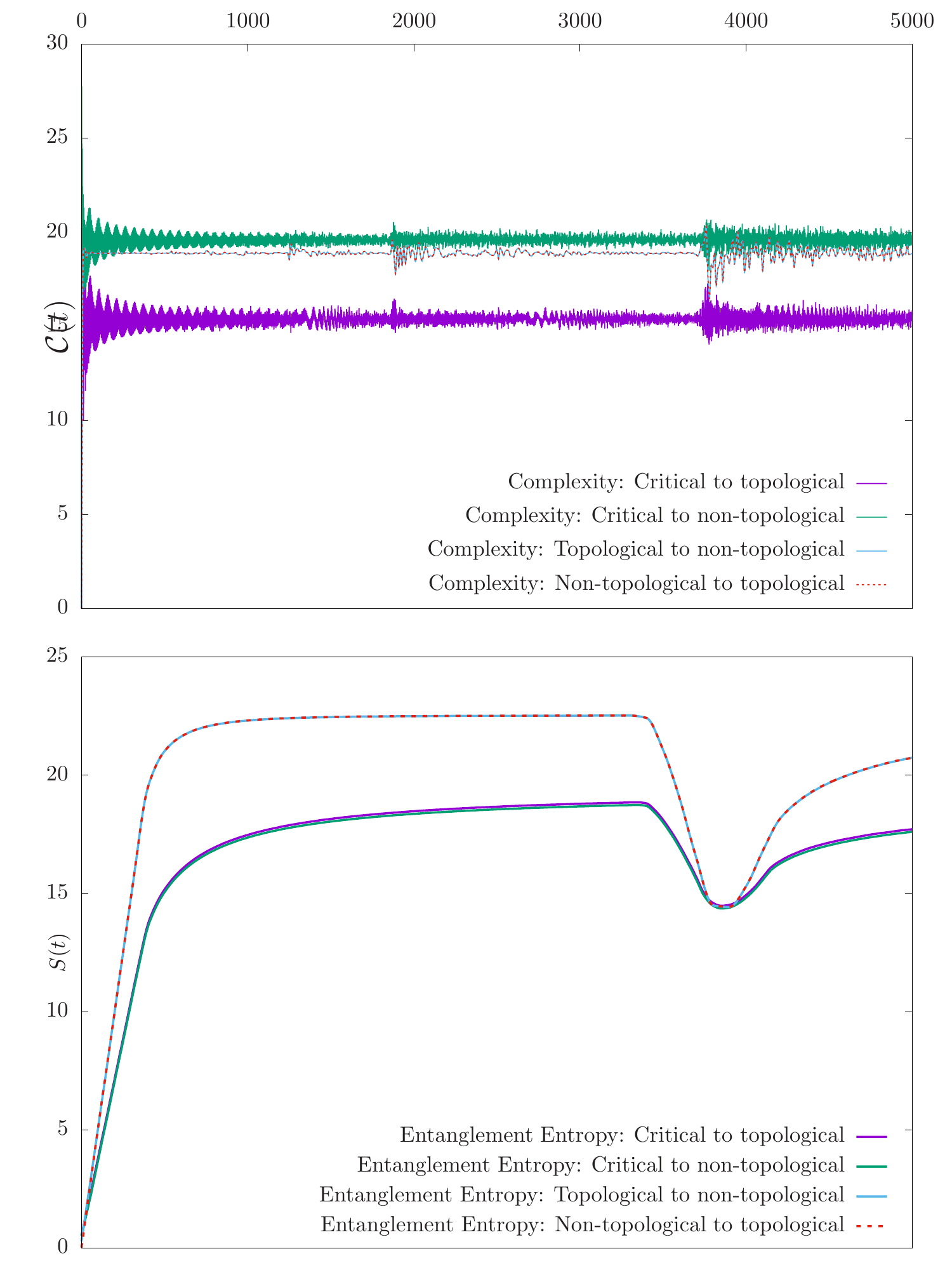}
	\caption{The complexity from the Fubini-Study line element for $N$=1500. 
		Quenching 
		(a) from a massive phase to the critical point
		(b) to a massive phase.
	}
	\label{fig:FScomplexity}
\end{figure}

\begin{figure}[h!]
	\centering
	\includegraphics{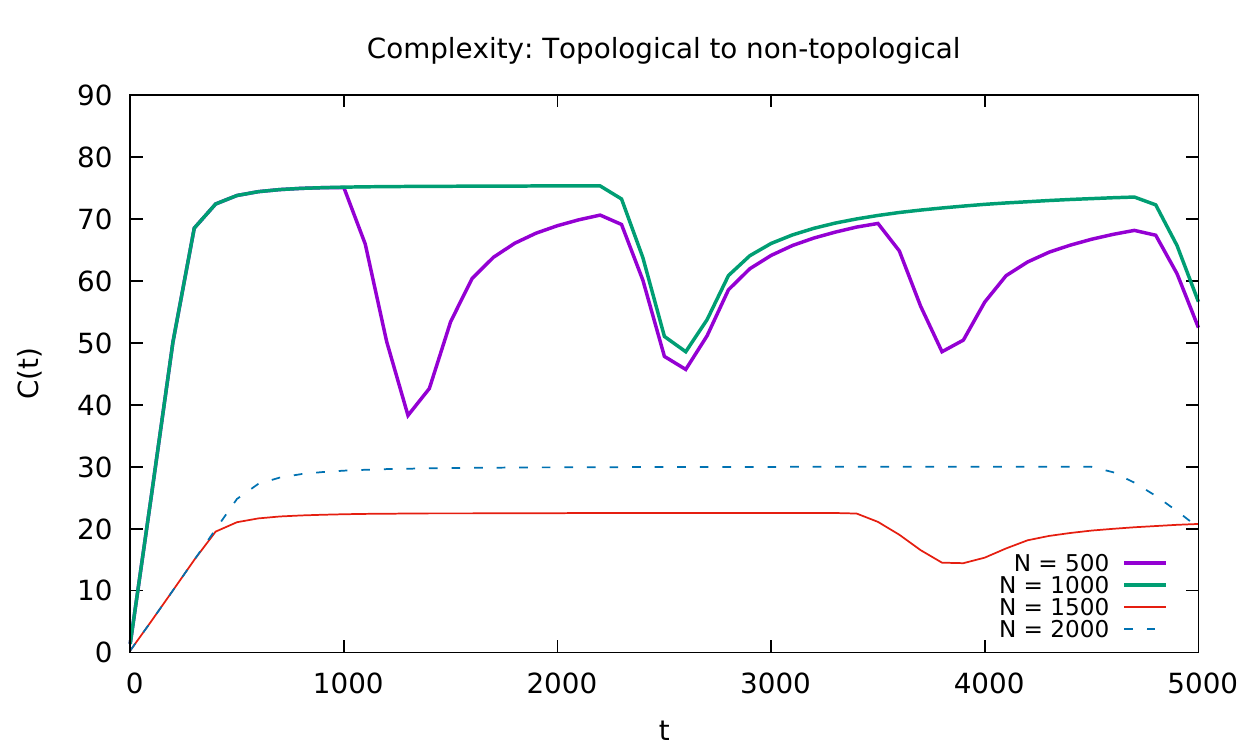}
	\caption{The circuit complexity for different system sizes $N$}
	\label{fig:circuitcomplexityN}
\end{figure}

The evolution of the complexity is shown in Figs.~(\ref{fig:circuitcomplexity}) and (\ref{fig:FScomplexity}) --
Fig.~(\ref{fig:circuitcomplexity}) shows results for the circuit complexity, while 
Fig.~(\ref{fig:FScomplexity}) shows results from the Fubini-Study line element; in both figures, 
panel (a) shows quenches from the massive phases to the QCP, while 
panel (b) shows quenches to a massive phase.
For all the different quenches, the complexity grows extremely rapidly and then saturates, 
with some oscillations about the saturation value.
When quenching to the QCP, distinct revivals appear in the complexity, similar to what occurs in the
EE. When quenching to a massive phase, revivals also occur, but they are more pronounced when quenching
between massive phases.
To better understand these results,
the insets of Fig.~(\ref{fig:circuitcomplexity}) show results for the 
negative logarithm of the fidelity: $-\ln {\cal F}_{12}$.
This quantity behaves similarly to the complexity; this gives insight into the behaviour of the complexity and, 
in particular, its rapid growth and then saturation -- this occurs because the target state becomes orthogonal 
to the reference state rapidly.
Note also that revivals appear in the fidelity when quenching to the QCP \cite{echo1,appeasement,echo2};
however, the signature in the complexity is more pronounced. We also note that the high-frequency oscillation observed in the complexity is an artefact of the finite system size $N$. Increasing the system size means that the complexity oscillates more slowly, as can be seen in Fig. (\ref{fig:circuitcomplexityN}) where we compare the same complexity at multiple values of $N$. Although we have shown for one particular case, this is true for all the cases. 

In both the circuit complexity and Fubini-Study approaches, the results are identical for the non-TI to TI 
and the TI to non-TI quenches -- the complexity is the shortest distance between states; 
the direction of the quench does not change this distance.
However, the Fubini-Study approach can distinguish between the non-TI $\leftrightarrow$ QCP and 
TI $\leftrightarrow$ QCP quenches, while the circuit complexity cannot;
the Fubini-Study can detect the topological order.
More generally, for this system the Fubini-Study approach provides a more sensitive measure of complexity 
than the circuit complexity; this is, in fact, contrary to what is expected \cite{brownsusskind,heller}.


\section{Discussion}

We investigated the evolution of complexity and entanglement 
following a quantum quench in  the SSH model.
Interestingly, for our system, the complexity saturated much more rapidly than the EE; this is contrary to 
what happens in the blackhole scenario, where the complexity is expected to saturate 
exponentially slower than the EE \cite{susskind2}.
We observed that the mechanism for this rapid saturation of the complexity is tied to the rapid decay of the fidelity, namely because the final state becomes orthogonal to the initial state very rapidly. The magnitude of the fidelity's decay is set by the parameters of the Hamiltonian e.g. if one starts in the topological phase with $q=0.2,q'=1$ and one quenches to a state still in the topological phase with, say $q=0.25,q'=1$, the magnitude of the fidelity's decay will be small.  While this is also expected to influence the rate of the decay, we do not have a quantitative understanding as to how this occurs. 
In future work, we intend to pursue the precise relationship between the decay of the fidelity and the saturation of the complexity. 

Another key observation was of revivals in the complexity. Note that the complexity is (much) less demanding 
computationally than measures of entanglement, yet the complexity, particularly the Fubini-Study approach, captures 
similar information such as revivals and information about different topological transitions. 
This makes the complexity rather appealing as a probe of quantum phase transitions.

Our investigations revealed a shortcoming of complexity in systems with topological order (TO). 
The TO of the SSH model derives from the nontrivial topology of its Hilbert 
space \cite{TIreviews}. TOs are readily probed on a manifold with a boundary -- 
gapless edge modes arise due to the TO \cite{wenbook}. 
As an entanglement partition behaves as a physical boundary \cite{srednicki}, gapless modes 
arise in systems with TO, with support along the partition.
We saw how the TO shows itself distinctly in the entanglement -- the EE was always larger in the 
TI phase (due to the edge modes); the entanglement spectrum provides a ``smoking gun" for the TO, as the low-energy, 
universal part of the entanglement spectrum is due to these edge modes \cite{haldane}.
The complexity, on the other hand, was not as sensitive a probe of TO -- 
the Fubini-Study approach could sense that the TI and non-TI phases were different
(as it could distinguish between the non-TI $\leftrightarrow$ QCP and 
TI $\leftrightarrow$ QCP quenches). The circuit complexity, on the other hand, could
not differentiate the two phases.
Interestingly, our results showed the Fubini-Study approach to be a more sensitive measure of complexity
for this system (than the circuit complexity), contrary to what is expected \cite{brownsusskind,heller}.


Finally, this work takes steps toward bringing complexity closer to physical systems and, 
in particular, experiment.
Recently, measuring entanglement has been an extremely active/exciting avenue of research \cite{measureentangle}.
As entanglement has been measured in cold atom systems \cite{measurecoldatom1,measurecoldatom2,
measurecoldatom3} (where the SSH model has been realized), it is not unreasonable to expect some of 
the ideas discussed here could be probed experimentally.


\section*{Acknowledgements} 
AB thanks Aninda Sinha, Pratik Nandy, and Tadashi Takayanagi for useful discussions. 
 A.B. is supported by Research Initiation Grant (RIG/0300) provided by IIT-Gandhinagar and Start-Up Research Grant (SRG/2020/001380) by Department of Science \& Technology Science and Engineering Research Board (India). 
EHK acknowledges support from funds from the University of Windsor.
NM is supported by the South African Research Chairs Initiative of the Department of Science and Technology 
and the National Research Foundation (NRF) of South Africa; 
he was supported, in part, by the Perimeter Institute for Theoretical Physics during the course of this work.
Research at the Perimeter Institute is supported by the Government of Canada through the 
Department of Innovation, Science, and Economic Development, and by the 
Province of Ontario through the Ministry of Research and Innovation. 
Any opinion, finding and conclusion or recommendation expressed in this material is that of the authors and 
the NRF does not accept any liability in this regard.

\begin{appendix}
\section{Details of the System }\label{app}
\subsection{The System} \label{secc3}

We consider a fermionic Hamiltonian of the form
\begin{equation}
 \hat{H}_F = \sum_k \Psi^{\dagger}_k
     \left( \begin{array}{cc}
       \xi^{\phantom *}_k \ &  \Delta^{\phantom *}_k  \\
       \Delta^*_k \ &  -\xi^{\phantom *}_k  
     \end{array} \right)  
  \Psi^{\phantom \dagger}_k \ ,
\label{Hfermion}
\end{equation}
where $\Psi_k$ is the spinor
\begin{equation}
  \Psi^T_k = \left( \tilde{a}_k , \tilde{b}_k \right)  \ .
\end{equation}
[For the SSH model,  $\xi_k = 0$, $\Delta_k = q+ q' \exp(-ik)$.]
Eq.~(\ref{Hfermion}) is readily diagonalized
\begin{equation}
 \hat{H}_F = \sum_k E_k \left( 
     \alpha^{\dagger}_k \alpha^{\phantom \dagger}_k
  - \beta^{\dagger}_k \beta^{\phantom \dagger}_k 
  \right) \ ,
\end{equation}
where $E_k = \sqrt{ \xi_k^2 + |\Delta^{\phantom 2}_k|^2 }$, and the
$\{ \alpha_k, \beta_k \}$ are related to the $\{ \tilde{a}_k, \tilde{b}_k \}$ 
by the Bogoliubov transformation
\begin{subequations}
\begin{equation}
     \left( \begin{array}{c}
      \tilde{a}_k  \\  \tilde{b}_k  
     \end{array} \right) 
   =     \left( \begin{array}{cc}
      u^{\phantom *}_k \ & -v^*_k \\
      v^{\phantom *}_k \ & u^*_k
     \end{array} \right) 
      \left( \begin{array}{c}
      \alpha_k \\  \beta_k 
     \end{array} \right)  
\label{bogoliubov}
\end{equation}
with
\begin{equation}
 u_k = \sqrt{ \frac{1}{2} \left( 1 + \frac{\xi_k}{E_k} \right) }
 \ \ , \ \ 
 v_k = \frac{\Delta_k^*}{|\Delta_k|} \sqrt{ \frac{1}{2} \left( 1 - \frac{\xi_k}{E_k} \right) }  \ .
\end{equation}
\end{subequations}
%
Physically, $\alpha_k$ ($\beta_k$) destroys a fermion of momentum-$k$ 
in the conduction (valence) band.


\subsection{Quantum Quench} \label{secc}


\subsubsection{Quench Protocol}

We consider a quench in the above model ---
\begin{subequations}
\begin{equation}
 \underline{t<0}: \hat{H} = \hat{H}_< (q^{\phantom '}_<,q'_<)
 \ \  ,  \ \
 \underline{t>0}: \hat{H} = \hat{H}_> (q^{\phantom '}_>,q'_>)
 \ .
\end{equation}
More explicitly, we consider 
\begin{equation}
 \underline{t < 0}: 
 \hat{H}_< = \sum_k \Psi^{\dagger}_k
    \left( \begin{array}{cc}
      \xi_{k}^< \ &  \Delta_{k}^<  \\
      (\Delta_{k}^<)^* \ &  -\xi_{k}^<  
     \end{array} \right)  \Psi^{\phantom \dagger}_k
 \ \ ; \ \ 
 \underline{t > 0}: 
 \hat{H}_> = \sum_k \Psi^{\dagger}_k
    \left( \begin{array}{cc}
      \xi_{k}^> \ &  \Delta_{k}^>  \\
      (\Delta_{k}^>)^* \ &  -\xi_{k}^>  
     \end{array} \right)  \Psi^{\phantom \dagger}_k  \ .
\label{quenchfermion}
\end{equation}
\end{subequations}
Let $\{\alpha_{<k},\beta_{<k} \}$ ($\{\alpha_{>k},\beta_{>k} \}$)
be eigenoperators of $\hat{H}_<$ ($\hat{H}_>$):
\begin{subequations}
\begin{eqnarray}
     \left( \begin{array}{c}
      \tilde{a}_k  \\  \tilde{b}_k  
     \end{array} \right) 
   =     \left( \begin{array}{cc}
      u^{\phantom *}_{<k} \ & -v^*_{<k} \\
      v^{\phantom *}_{<k} \ & u^*_{<k}
     \end{array} \right) 
      \left( \begin{array}{c}
      \alpha_{<k} \\  \beta_{<k} 
     \end{array} \right)  
   & \longrightarrow & 
    \hat{H}_< = \sum_k E_k^< \left( 
     \alpha^{\dagger}_{<k} \alpha^{\phantom \dagger}_{<k}
  - \beta^{\dagger}_{<k} \beta^{\phantom \dagger}_{<k} 
  \right) \ ,
 \\
     \left( \begin{array}{c}
      \tilde{a}_k  \\  \tilde{b}_k  
     \end{array} \right) 
   =     \left( \begin{array}{cc}
      u^{\phantom *}_{>k} \ & -v^*_{>k} \\
      v^{\phantom *}_{>k} \ & u^*_{>k}
     \end{array} \right) 
      \left( \begin{array}{c}
      \alpha_{>k} \\  \beta_{>k} 
     \end{array} \right)  
    & \longrightarrow & 
    \hat{H}_> = \sum_k E_k^> \left( 
     \alpha^{\dagger}_{>k} \alpha^{\phantom \dagger}_{>k}
  - \beta^{\dagger}_{>k} \beta^{\phantom \dagger}_{>k} 
  \right) \ .
\end{eqnarray}
\end{subequations}
The  $\{\alpha_{>k},\beta_{>k} \}$ are related to the 
$\{\alpha_{<k},\beta_{<k} \}$ by the unitary transformation
\begin{subequations}
\begin{equation}
      \left( \begin{array}{c}
      \alpha_{>k}  \\  \beta_{>k}  
     \end{array} \right) 
   =     \left( \begin{array}{cc}
      {\cal U}^{\phantom *}_k \ & -{\cal V}^*_k \\
      {\cal V}^{\phantom *}_k \ &  {\cal U}^*_k
     \end{array} \right) 
      \left( \begin{array}{c}
      \alpha_{<k} \\  \beta_{<k} 
     \end{array} \right)
\end{equation}
where
\begin{equation} 
  {\cal U}_k = u^*_{>k} u^{\phantom *}_{<k} 
                  + v^*_{>k} v^{\phantom *}_{<k}
 \ \ , \ \ 
  {\cal V}_k = u^*_{>k} v^{\phantom *}_{<k} 
                   - v^*_{>k} u^{\phantom *}_{<k}
 \ .
\end{equation}
\end{subequations}
Then, $\hat{H}_>$ can be written in terms of the eigenoperators of $\hat{H}_<$ as
\begin{equation}
 \hat{H}_> = \sum_k 2E^>_k \left[
 \left( \mid {\cal U}_k \mid^2 -\mid {\cal V}_k \mid^2 \right) \tau_k^z
 - {\cal U}_k^* {\cal V}_k^* \tau_k^+ - {\cal U}_k {\cal V}_k \tau_k^- 
 \right]  \ ,
\end{equation}
 where 
\begin{equation}
 \tau_k^- = \beta^{\dagger}_{<k} \alpha^{\phantom \dagger}_{<k}
\ , \
 \tau_k^+ = \alpha^{\dagger}_{<k} \beta^{\phantom \dagger}_{<k}
\ , \
 \tau_k^z = ( \alpha^{\dagger}_{<k} \alpha^{\phantom \dagger}_{<k}
                  - \beta^{\dagger}_{<k} \beta^{\phantom \dagger}_{<k} ) / 2
\end{equation}
obey an $SU(2)$ algebra.


\subsection{Reference and Target States} \label{secc4}

The time evolution operator can be written as
\begin{subequations}
\begin{equation}
 \hat{U}(t) = \prod_k \exp\left( \omega^{\phantom *}_k \tau_k^z 
   + \alpha_k^+ \tau_k^+ + \alpha_k^- \tau_k^- \right)  \ ,
\label{evolution}
\end{equation}
where
\begin{equation} 
 \{ \omega^{\phantom *}_k = -it 2 E_k^> \left( \mid {\cal U}_k \mid^2 -\mid {\cal V}_k \mid^2 \right) 
  \ , \  
     \alpha_k^+ = it 2E_k^>~ {\cal U}_k^* {\cal V}_k^* 
  \ ,\  
     \alpha_k^- = it 2 E_k^>~ {\cal U}_k {\cal V}_k  \}  \ .
\end{equation}
\end{subequations}
In what follows, we take as our reference state, $| \psi_R \rangle$, 
the ground state of $\hat{H}_<$;
we take as our target state, $| \psi_T \rangle$, the state obtained by evolution with $\hat{H}_>$:
$| \psi_T \rangle = \exp(-i t \hat{H}_>) | \psi_R \rangle$.
The initial state has the valence band of $\hat{H}_<$ completely filled 
\begin{subequations}
\begin{equation}
 | \psi_R \rangle = \prod_k \beta^{\dagger}_{<k} | 0 \rangle  \ ,
\label{finitial}
\end{equation}
where $| 0 \rangle$ is the Fock vacuum. The final state is given by
\begin{equation}
 | \psi_T \rangle = \prod_k \frac{1}{\sqrt{1 + | z_k(t) |^2 } }
   \left( \beta^{\dagger}_{<k} + z_k(t) \alpha^{\dagger}_{<k} \right) 
   | 0 \rangle  \ ,
\label{ffinal}
\end{equation}
where
\begin{equation}
  z_k(t) = \frac{i 2~ {\cal U}_k^* {\cal V}_k^* \sin(E_k^> t) }{
       \cos(E_k^> t) + i \left( \mid {\cal U}_k \mid^2 - \mid {\cal V}_k \mid^2 \right) 
       \sin(E_k^> t) }  \ .
\label{zexplicit}
\end{equation}
\end{subequations}
Note: 
To obtain Eqs.~(\ref{ffinal}) and (\ref{zexplicit}), we utilized the decomposition \cite{perelomovsupp}
\begin{subequations}
\begin{equation}
   \exp\left( \omega \tau^z + \alpha^+ \tau^+ + \alpha^- \tau^- \right)
 = \exp(\gamma_+ \tau^+) \exp\left[ (\ln \gamma_0 ) \tau^z \right]
    \exp(\gamma_- \tau^-) \ ,
\end{equation}
where
\begin{equation}
 \gamma^0 = \left[ \cosh \Omega - \frac{\omega}{2\Omega} \sinh \Omega \right]^{-2}
 \ , \
 \gamma^{\pm} = \left( \frac{\alpha^{\pm}}{2\Omega} \right)
  \frac{\sinh \Omega}{\cosh \Omega - (\omega/2\Omega) \sinh \Omega} \ ,
\end{equation}
with
\begin{equation}
   \Omega^2 = \omega^2/4 + \alpha^+ \alpha^-  \ .
\end{equation}
\end{subequations}
[To simplify the notation, we have written $\gamma^+$ as $z$ in Eq.~(\ref{ffinal}).]


\section{State Manifold: Riemannian Metric; Geodesics and Complexity}\label{app2}

The space of states obtained by evolution with Eq.~(\ref{evolution}) can be given a 
Riemannian structure \cite{statemanifold}. Starting with the Bures distance
\begin{equation}
 D^2_{12} = 2 \left[ 1 - {\cal F}_{12} \right]
\nonumber
\end{equation}
where ${\cal F}_{12}$ is the fidelity ${\cal F}_{12} = | \langle \psi_1 | \psi_2 \rangle |$,
one considers two nearby states -- Tayor expanding gives the line element \cite{statemanifold}
\begin{subequations}
\begin{equation}
  ds^2 = \sum_{ij} g_{ij}~ dx^i dx^j  \ ,
\end{equation}
where $ds^2 \equiv {\rm lim}_{| \psi_2 \rangle \rightarrow | \psi_1 \rangle} D^2_{12}$,
and we have introduced the metric tensor
\begin{equation}
  g_{ij} = \rm{Re} \left[ \langle \partial_i \psi \mid \partial_j \psi \rangle \right]
    - \langle \partial_i \psi \mid \psi \rangle \langle \psi \mid \partial_j \psi \rangle
 \ .
\label{metric}
\end{equation}
\end{subequations}

For states of the form Eq.~(\ref{ffinal}), the fidelity is given by
\begin{equation} 
   {\cal F}_{12} = \prod_k \frac{ | 1 + z^{*}_{1k} z^{\phantom *}_{2k} | }
   { \sqrt{1 + |z_{1k}|^2} \sqrt{1 + |z_{2k}|^2 } };
\end{equation}
one obtains the line element
\begin{subequations}
\begin{equation}
  ds^2 = \sum_k \frac{ |dz_k|^2}{ \left( 1 + |z_k|^2 \right)^2}  \ .
\label{appendixCP1}
\end{equation}
$\forall~k$ this is the line element for a two-sphere $S^2$ in the $CP^1$ representation.
Writing $z_k = |z_k| \exp(i\phi_k)$ with $|z_k| = \tan \theta_k/2$,
one obtains the line element in the spherical coordinates
\begin{equation}
   ds^2 = \sum_k \frac{1}{4} \left( d\theta_k^2 + \sin^2\theta_k~ d\phi_k^2 \right)  \ .
\label{appendixspherical}
\end{equation}
\end{subequations}
%
This result (Eqs.~(\ref{appendixCP1}) and (\ref{appendixspherical})) is already anticipated by 
Eq.~(\ref{evolution}) -- the time-evolution operator moves the system on the group manifold; 
$\forall \ k$, Eq.~\ref{evolution} moves the system on the Bloch sphere.

For a particular momentum, $k$, the arc-length between two points, $s_k$, is
\begin{equation}
 s_k = \int_{0}^{t} \hspace{-0.07in} dt'~ 
  \frac{1}{1 + |z_k|^2 }  \left| \frac{dz_k}{dt'} \right|  \ .
\end{equation}
The geodesics are great circles; the arc-length of the geodesic, ${\cal C}_k$, is
\begin{subequations}
\begin{equation}
 {\cal C}_k = \frac{1}{2} \arccos\left[ \cos \theta_1 \cos \theta_2 
   + \sin \theta_1 \sin \theta_2 \cos (\phi_1 - \phi_2) \right]  \ ,
\end{equation}
where $(\theta_1,\phi_1)$ ($(\theta_2,\phi_2)$) are the spherical coordinates 
of the initial (final) point.  [The prefactor of $1/2$ is the radius of the sphere.]
As the initial state is given by Eq.~(\ref{finitial}), one obtains 
($|z_k| = \tan \theta_k/2$)
\begin{equation}
 {\cal C}_k = \arctan |z_k|  \ .
\end{equation}
\end{subequations}

For the full state manifold, $S^2 \times S^2 \times \cdots \times S^2$, define a metric, $s$, 
as \cite{metricspaces}
\begin{equation}
  s = \sqrt{ \sum_k s_k^2 }  \ ;
\end{equation}
it follows 
the system's complexity, ${\cal C}$, is
\begin{equation}
 {\cal C} = \sqrt{ \sum_k {\cal C}_k^2 }  \ .
\end{equation}


\section{Complexity and Entanglement from the Correlation Matrix}\label{app1}

Define the {\sl correlation matrix} $\hat{C}_k$ \cite{peschel,ehk} -- 
\begin{equation}
 \hat{C}_{k}(t) = \langle \psi(t) | \Psi^{\phantom \dagger}_{k} 
                       \Psi^{\dagger}_{k} | \psi(t) \rangle \ ,
\label{Cmomentum}
\end{equation}
where $\Psi_{k}^T = ( \tilde{a}_{k}, \tilde{b}_{k} )$. 
Explicitly, one obtains
\begin{equation}
 \hat{C}_{k}(t) 
  =  \left( \begin{array}{cc}
       |u_k(t)|^2 \ &  u_k(t) v^*_k(t)  \\
       v_k(t) u^*_k(t) \ &  |v_k(t)|^2  
     \end{array} \right)  \ ,
\end{equation}
where
\begin{equation}
    \left( \begin{array}{c}
      u_{k}(t)  \\  v_{k}(t)  
     \end{array} \right) 
   =  \exp(-i t H_{>} )
      \left( \begin{array}{c}
      u_{<k} \\  v_{<k} 
     \end{array} \right)
 \ \ {\rm with} \ \ 
 H_{>}  =  \left( \begin{array}{cc}
      \xi_{k}^> \ &  \Delta_{k}^>  \\
      (\Delta_{k}^>)^* \ &  -\xi_{k}^>  
     \end{array} \right)  \ .
\end{equation}
From $\hat{C}_k(t)$, one obtains the complexity and measures of entanglement.


\subsection{Complexity} \label{secc1}

We consider evolving the initial correlation matrix $\hat{C}_k$ by the unitary operator $U_k$, 
\begin{equation}
 \tilde{C}_k(s) \equiv U_k^{\phantom \dagger} (s) 
                       \hat{C}_k(t=0) U_k^{\dagger}(s) \ ,
\end{equation}
with the boundary condition $\tilde{C}_k(s=1) = \hat{C}_k(t)$
i.e. at $s=1$ we recover the final correlation matrix.
We parameterize $U_k$ as the path-ordered exponential
\begin{equation}
 U_k(s) = {\cal P} \exp \left[ - i \sum_I \int_0^1 \hspace{-0.07in} ds~
              Y_k^I(s) M_I \right]  \ ,
\label{gates}
\end{equation}
where the $\{ M_I \}$ are group generators, the $\{ Y_k^I \}$ are control functions,
and $s$ parameterizes the path;
we define a ``cost functional" on the space of unitaries ${\cal C}[\{ U_k \}]$ via
\begin{equation}
 {\cal C}[ \{ U_k \} ] = \int_0^1 \hspace{-0.07in} ds
    \sqrt{ \sum_k \sum_I  |Y_k^I(s)|^2  } \ .
\label{cost}
\end{equation}
 Our goal is to minimize this function w.r.t $Y^{I}(s)$ as a function of $s$. This will  determine the optimal path, 
namely the geodesic (with respect to the cost functional) on the space of unitaries; 
the complexity is the arc-length of the geodesic.

Taking the $\{ M_I \}$ to be generators (in the fundamental representation) of SU(2), 
\begin{equation} 
\left\{
 M_1 =  \left( \begin{array}{cc}
       0 \ & 1  \\
       1 \ & 0  
     \end{array} \right) 
 \ , \ 
 M_2 =  \left( \begin{array}{cc}
       0 \ & -i  \\
        i \ & 0  
     \end{array} \right) 
 \ , \ 
 M_3 =  \left( \begin{array}{cc}
       1 \ & 0  \\
       0 \ & -1  
     \end{array} \right) 
 \ , \
 M_0 =  \left( \begin{array}{cc}
       1 \ & 0  \\
       0 \ & 1  
     \end{array} \right)   
\right\}  \ ,
\nonumber
\end{equation}
$U_k$ is readily parametrized as
\begin{subequations}
\begin{equation}
 U_k(s) =  \left( \begin{array}{cc}
       \cos \rho_k(s) \exp [ i \phi_k(s) ] \ & -i \sin \rho_k(s) \exp[-i \chi_k(s) ]  \\
       -i \sin \rho_k(s) \exp[ i \chi_k(s) ] \ &  \cos \rho_k(s) \exp[ -i \phi_k(s) ]  
     \end{array} \right)  \ .
\end{equation}
Then, using that $U_k(s)$ satisfies the Schrodinger-like equation
\begin{equation}
 i \partial_s U_k(s) = \sum_I Y^I_k(s) M_I~ U_k(s) \ ,
\nonumber
\end{equation}
one extracts the $\{ Y^{I}_k \}$ as 
\begin{eqnarray}
 Y^{1}_k = -i \left[ \cos \beta_k~ \frac{d\rho_k}{ds} 
        + \frac{1}{2} \sin 2\rho_k \sin \beta_k \left( \frac{d\phi_k}{ds} 
                         + \frac{d\chi_k}{ds} \right)  \right],\,
                          \nonumber \\ 
 Y^{2}_k = - \sin \beta_k~ \frac{d\rho_k}{ds}  
        + \frac{1}{2} \sin 2\rho_k \cos \beta_k \left( \frac{d\phi_k}{ds} 
                         + \frac{d\chi_k}{ds} \right)  \ ,
 \nonumber \\ 
 Y^{3}_k = - \cos^2 \rho_k~ \frac{d\phi_k}{ds} + \sin^2 \rho_k \frac{d\chi_k}{ds}\,,\nonumber\\
 Y^{0}_k = 0  \ ,
\end{eqnarray}
where $\beta_k = \phi_k - \chi_k$;
from Eq.~(\ref{cost}), one obtains the (right-invariant) metric 
\begin{equation}
 ds^2 = \sum_k \left[ 
  d\rho_k^2 + \cos^2 \rho_k~ d\phi_k^2 +\sin^2 \rho_k~ d\chi_k^2  \right]  \ .
\end{equation}
\end{subequations}

We solve for the geodesic with the boundary conditions
\begin{subequations}
\begin{eqnarray}
 & & \rho_k(s=0) = 0 \ \  , \ \   \beta_k(s=0) = \chi_k^0 \ ,
  \\
 & & \rho_k(s=1) 
    = \frac{1}{2} \arccos \left[ \hat{C}_k(t)_{22} - \hat{C}_k(t)_{11} \right], \ \ 
 \beta_k(s=1) = \arctan \left[ - \frac{ \rm{Re}\hat{C}_k(t)_{12} }
                       { \rm{Im} \hat{C}_k(t)_{12} } \right], 
\end{eqnarray}
\end{subequations}
with $\chi_k^0$ arbitrary, and
the subscripts (11,22,12) denote the components of $\hat{C}_k$. 
A solution for the geodesic is \footnote{We can solve for $\phi_k$ or $\chi_k$ unambiguously, but not both of them.}
\begin{equation}
 \rho_k(s) = \rho_k(s=1) s , \ \ 
 \chi_k(s) = \chi_k^0
\ \ ,  \ \
 \phi_k(s) = 0  \ ;
\end{equation}
then the complexity is 
\begin{equation}
 {\cal C}[ \{ U_k \} ] = \frac{1}{2} \sqrt{ \sum_k 
     \left( \arccos \left[ \sin^2 (E_k^> t) \cos( 2\theta_k^> - 2 \theta_k^< )
  + \cos^2 (E_k^> t) \right] \right)^2 } \ .
\end{equation}


\subsection{Entanglement} \label{secc2}

We are interested in the entanglement of a spatial partition.
We partition the system into subsystems $A$ and $B$; measures of entanglement are obtained from 
the reduced density matrix for subsystem-$A$, $\hat{\rho}_A = {\rm Tr}[ \hat{\rho} ]$, with 
$\hat{\rho}$ being the density matrix of the entire system. 
For the system considered in this work, $\hat{\rho}_A$ has the form \cite{peschel,ehk}
\begin{subequations}
\begin{equation}
 \hat{\rho}_A = \frac{1}{\cal Z} \exp\left( - {\cal H}_A \right) \ ,
\end{equation}
where
\begin{equation}
 {\cal H}_A = \sum_{m,n \in A} 
  \Psi^{\dagger}_m H^A_{mn} \Psi^{\phantom \dagger}_n  \ ,
\end{equation}
\end{subequations}
and  $\Psi_{m}^T = ( a_{m}, b_{m} )$ is the spinor at site-$m$.
Measures of entanglement are readily obtained from the correlation matrix restricted to subsytem-$A$  \cite{peschel,ehk}
-- its elements are given by ($m,n \in A$)
\begin{equation}
 C_{mn}(t) = \langle \psi(t) | \Psi^{\phantom \dagger}_m \Psi^{\dagger}_n | \psi(t) \rangle \ ;
\end{equation}
$\hat{C}_{mn}(t)$ can be obtained from $\hat{C}_k(t)$ via
\begin{equation}
 \hat{C}_{mn}(t) = \frac{1}{N} \sum_k \exp[i k (m-n)] \hat{C}_k(t) \ .
\end{equation}
Then, denoting the eigenvalues of $\hat{\bf C}$ by $\{ \lambda_n \}$
($0 \leq \lambda_n \leq 1$), the spectrum of $H^A$, $\{ E_n \}$, is obtained via
\begin{equation}
 E_n = \ln \left[ (1-\lambda_n)/\lambda_n \right]  \ .
\label{ES}
\end{equation}

A central measure of entanglement is the Renyi entropy
\begin{subequations}
\begin{equation}
 S_{\alpha} = \frac{1}{1-\alpha} 
  \ln {\rm Tr} \left[ \hat{\rho}_A^{\alpha} \right]  \ .
\end{equation}
Explicitly,
\begin{eqnarray}
 S_{\alpha} & = & \frac{1}{1-\alpha} \sum_n \left[
   \ln \left( 1 + e^{-\alpha E_n} \right) 
 - \ln \left( 1 + e^{-E_n} \right)^{\alpha} \right]  \ ;
\end{eqnarray}
using Eq.~(\ref{ES}), 
\begin{equation}
 S_{\alpha} = \frac{1}{1-\alpha} \sum_n 
 \ln \left[\lambda_n^{\alpha} + (1 - \lambda_n)^{\alpha} \right]  \ .
\end{equation}
\end{subequations}
Taking $\alpha \rightarrow 1^+$ gives the entanglement entropy
\begin{equation}
 S = - \sum \left[ \lambda_n \ln \lambda_n
   + (1 - \lambda_n) \ln ( 1 - \lambda_n ) \right]  \ .
\end{equation}
\end{appendix}

\end{document}